\documentclass[%
article 
 amsmath,amssymb,
 aps, physrev, showkeys,
]{revtex4-2}

\usepackage{graphicx}
\usepackage{dcolumn}
\usepackage{bm}
\usepackage{hyperref}
\usepackage[mathlines]{lineno}

\usepackage[dvipsnames]{xcolor}
\usepackage{url}
\usepackage{xcolor}
\usepackage{amsmath}
\usepackage[normalem]{ulem}
\usepackage{multirow}
\hypersetup{colorlinks=true, linkcolor=blue, filecolor=magenta, urlcolor=cyan}
\usepackage{orcidlink}
\usepackage{rotating}

\begin{document}

\title{\textbf{ALMA view on the nature of the compact VLA continuum sources \\ in the massive young stellar object G25.65+1.05}}

\author{N.~N.~Shakhvorostova \orcidlink{0000-0002-6180-9474}}
\email{Contact author: nadya@asc.rssi.ru}
 \affiliation{P.N.~Lebedev Physical Institute of RAS, Moscow 119991, Russia}
 \affiliation{Kotel'nikov Institute of Radio Engineering and Electronics of RAS, Moscow 125009, Russia}
\author{A.~M.~Sobolev \orcidlink{0000-0001-7575-5254}}%
 \affiliation{Xinjiang Astronomical Observatory, Chinese Academy of Sciences, Urumqi 830011, PR China }
\author{D.~A.~Ladeyshchikov \orcidlink{0000-0002-3773-7116}}%
 \affiliation{Ural Federal University, Yekaterinburg 620062, Russia}
  \author{S.~Y. Parfenov}%
 \affiliation{Ural Federal University, Yekaterinburg 620062, Russia}
\author{A.~A. Shagabutdinov \orcidlink{0000-0003-3030-0998}}%
 \affiliation{Ural Federal University, Yekaterinburg 620062, Russia}
  \author{Liu, Sheng-Yuan}%
 \affiliation{Institute of Astronomy and Astrophysics, Academia Sinica, Taipei 106319, Taiwan}


\begin{abstract}

This paper presents high-resolution ALMA observations of the massive young stellar object G25.65+1.05, which is known to host water maser super flares. To investigate the nature of compact continuum sources that have been previously identified in this region, we analyzed 1.3 mm dust continuum and molecular line emission. The central millimeter peak MM1 coincides with the centimeter source VLA~2, has a complex molecular spectrum, and is identified as a hot molecular core. Molecular emission of SiO and CH$_3$CN in the vicinity of MM1 reveals kinematics consistent with wide-angle outflow structures and a possible rotating disk in the source. VLA sources 1A, 1B, and 3 are lacking compact millimeter counterparts and trace shocked regions where the outflow interacts with the surrounding material. In particular, VLA~1A, the site of H$_2$O maser super flares, is interpreted as a shock interface that exhibits developed turbulent movements seen in the SiO molecular line. The observed turbulence creates conditions required for H$_2$O maser action, directly linking the nature of VLA~1A to the origin of the H$_2$O maser super flares.
\end{abstract}

\keywords{interstellar medium; masers; star-forming regions}

\maketitle

\section{Introduction}
\label{sec:intro}

Formation of massive stars is a complex astrophysical process, often obscured by dense natal environments \cite{2018ARA&A..56...41M}. Numerous phenomena accompany this process, such as protostellar disks, jets and outflows associated with young stellar objects, accretion bursts, shocks and cosmic masers (see, for example, \cite{2020SSRv..216...62R, 2020SSRv..216...43Z} and references therein). 
These rich and complex phenomena are primarily observed at radio wavelengths, as high-mass protostars are deeply embedded within and obscured by their natal gas clouds at optical wavelengths. Furthermore, the large distances to most high-mass star-forming regions (typically several kiloparsecs) hinder direct observation.

G25.65+1.05 emerges as a complex and highly active region of massive star formation, distinguished primarily by its extraordinary "super flare"\, 22 GHz H$_2$O maser activity. H$_2$O maser flares in G25.65+1.05 reached unprecedented peaks up to $\sim$76 kJy with rapid variability timescales \cite{2018ARep...62..213L, 2018ARep...62..584S, 2017ATel10728....1V, 2017ATel10853....1V, 2017ATel11042....1A} and, thus, became a target for multiple follow-up interferometric observations \cite{Bayandina2019, Burns2020, Bayandina2020, Bayandina2023, Shakhvorostova2025}. The high-resolution radio interferometry unraveled the region's structure and complexity at small scales, and the most probable physical mechanism lying behind the super flares was suggested. 
It was proposed that the sharp flares resulted from an increase in the optical depth of the maser medium, caused by the overlap of masering regions along the line of sight.

The high-resolution VLA observations \cite{Bayandina2023} resolved previously observed centimeter continuum emission into multiple components. Crucially, the source associated with the H$_2$O super-flares (VLA 1) was resolved into two distinct objects: VLA 1A and VLA 1B (see Fig.~1 in \cite{Bayandina2023}). VLA 1A and 1B represent peaks in the east-west elongated centimeter continuum source with associated H$_2$O maser emission, interpreted by \citet{Bayandina2023} as tracing a young protostellar jet. VLA 1A hosts the flaring maser spots arranged in a distinctive "V-shaped"\, morphology with the apex coinciding with the VLA 1A peak. The remaining maser spots are located between the VLA 1A and 1B peaks. A second continuum object, VLA 2, was identified as a massive protostar actively accreting and ejecting material. 
Fainter continuum sources VLA 3 and VLA 4, with flat spectral indices \cite{Bayandina2023}, have unclear nature and may represent more evolved objects or unrelated components and need further investigation. The dust continuum emission at 2.8 mm was previously observed in \cite{2000A&A...361.1095D} using the Plateau de Bure Interferometer with the angular resolution 4$^{''}\times$3$^{''}$, which was insufficient to reveal the fine spatial structure of the dust continuum.

The spatial and temporal characteristics of the H$_2$O maser flares near VLA 1A strongly favor a local amplification mechanism over excitation by the flare of the nearby source, such as the one that occurred in NGC~63334~I \cite{2018ApJ...866...87B, 2024A&A...691A.157V}. This was supported by the decrease of infrared activity in the G25.65+1.05 vicinity during the super flare period \cite{2019RAA....19...38S}. The rapid variability and the location of the flaring spot precisely at the apex of the "V-shaped"\, maser distribution point towards an increase in maser path length due to the chance line-of-sight superposition of distinct maser-emitting structures. This mechanism was theoretically investigated by \citet{1989ApJ...340L..17D} and proposed as an explanation of water maser super flares in Orion~KL \cite{2005ApJ...634..459S}. 

Despite these advances, the nature of the compact continuum sources associated with the H$_2$O maser super flares remains unclear. Why were H$_2$O maser super flares observed only in particular region VLA 1A and not in other parts of G25.65+1.05? Also, the distance to G25.65+1.05 is highly uncertain, with estimates ranging from $\sim$2.08 kpc (BeSSeL Survey probability; \citet{2016ApJ...823...77R}) to 12.5 kpc (H~I self-absorption; \citet{2011MNRAS.417.2500G}), critically impacting luminosity estimates for both the flares and the protostellar sources. 

This paper presents the results of high-sensitivity ALMA observations of the dust continuum and molecular lines in the high-mass star-forming region G25.65+1.05, building upon the foundation laid by previous works to further investigate the physics behind the extreme maser phenomena in this unique region. This work focuses on the structures and driving sources associated with the flaring H$_2$O maser emission, as probed by 1.3-mm dust continuum along with molecular lines indicative of shocks and dense gas.

\section{Observations and data reduction}
\label{sec:obs}

The 1.3 mm observations of G25.65+1.05 were carried out with the Atacama Large Millimeter/submillimeter Array (ALMA) in Cycle 8 on August 14, 2022, with the 12 m array in Band~6 (project 2021.1.00311.S, PI: Liu, Sheng-Yuan). The 12-m array configuration provided a maximum recoverable scale of $\sim$3.6$''$.
The on-source time was 121 seconds. The field of view and beam size were 25.86$''$ and 0.38$''$$\times$0.34$''$ at PA=7.7$^\circ$, respectively. The observation covered four spectral windows in Band 6: 216.63$-$218.50~GHz, 219.00$-$220.87~GHz, 230.04$-$231.92~GHz, and 231.86$-$233.74~GHz. 

Calibration, reduction, and imaging of the ALMA 12-m array data were done using the Common Astronomy and Software Applications (CASA) package (\url{http://casa.nrao.edu}), version 6.6.1-17. The continuum was subtracted using CASA task \texttt{uvcontsub} with the line-free channels selected automatically by CASA task \texttt{findcont}. In addition to a standard reduction, the self-calibration was performed automatically utilizing a new CASA task \texttt{hif\_selfcal} with default parameters. As a result of self-calibration, the continuum signal-to-noise ratio increased by a factor of 1.7. 
Data analysis and molecular line identification were performed using Cube Analysis and Rendering Tool for Astronomy (CARTA) \cite{2021zndo...4905459C} (\url{https://cartavis.org}).

\section{Results}
\label{sec:res}

\subsection{Continuum emission at 1.3 mm}

\begin{figure}
\begin{minipage}{0.49\textwidth}
\includegraphics[width=0.99\linewidth]{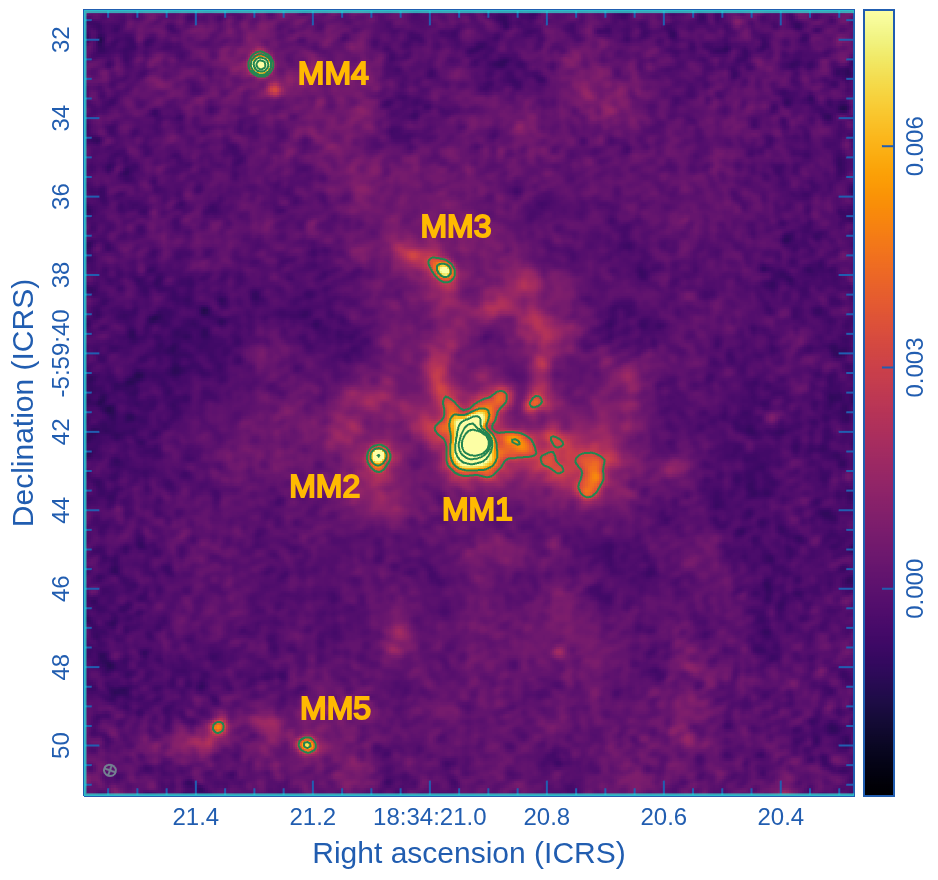}
\end{minipage}
\hfill
\begin{minipage}{0.49\textwidth}
\includegraphics[width=0.99\linewidth]{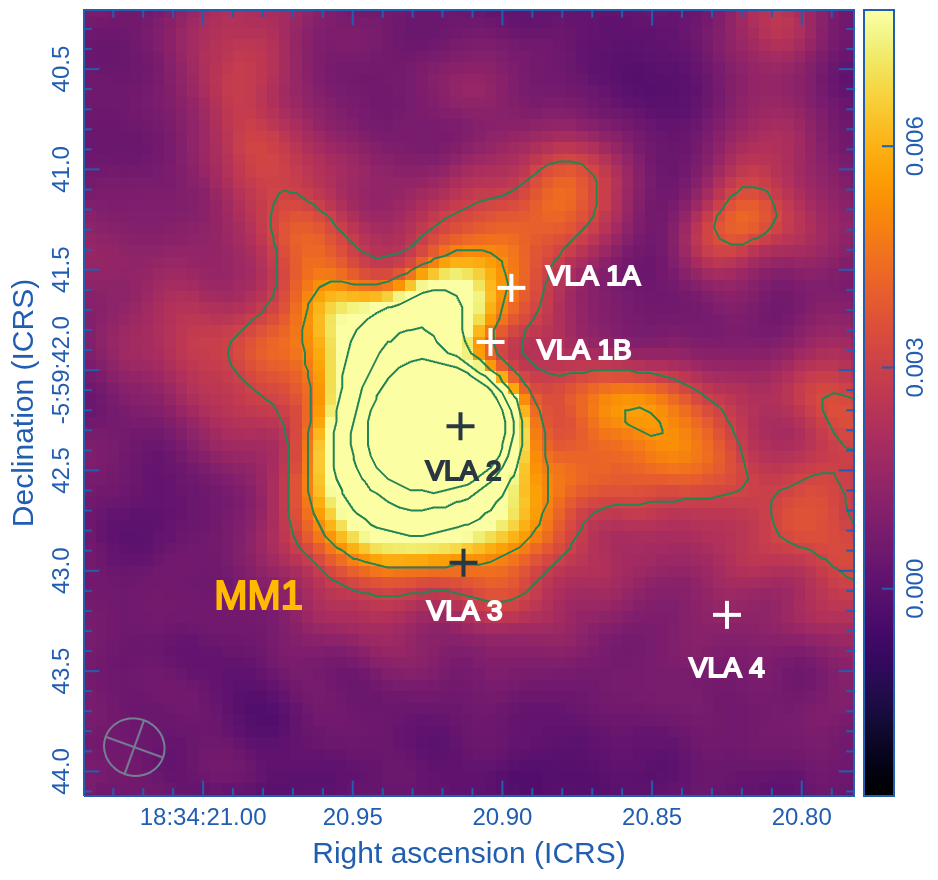}
\end{minipage}
\caption{{\bf Left panel}: Dust continuum emission at 1.3 mm in the G25.65+1.05 region observed with ALMA. The presented field size is 20$^{''}\times$20$^{''}$. The vertical colored panel indicates the brightness of emission in Jy/beam. The green contours are set to (3, 5, 9, 13, 17)$\times\sigma$, where $\sigma=1.003\times10^{-3}$~Jy/beam. {\bf Right panel}: Dust continuum emission at 1.3 mm in the G25.65+1.05 region observed with ALMA in the vicinity of the brightest millimeter continuum peak MM1. The vertical colored panel indicates the brightness of emission in Jy/beam. The green contours are set to (3, 5, 9, 13, 17)$\times\sigma$, where $\sigma=1.003\times10^{-3}$~Jy/beam. Positions of the continuum peaks VLA 1-4 detected at 15~GHz in the work \cite{Bayandina2023} are marked by crosses.}
\label{fig:cont}
\end{figure}

The high angular resolution ALMA observations of the G25.65+1.05 region at 1.3 mm reveal a complex dusty environment associated with high-mass star formation. The continuum emission map covering a field of $20^{''}\times20^{''}$ is presented in Fig.~\ref{fig:cont} at the left panel. The map shows a bright, central source surrounded by a network of fainter filaments and several compact condensations. The brightest 1.3~mm peak, designated MM1, is shown in Fig.~\ref{fig:cont} at the right panel and can be unambiguously identified as the central massive young stellar object (MYSO) in the region. This identification is confirmed by its precise spatial coincidence with the compact centimeter-wave source VLA~2 revealed at 15~GHz \cite{Bayandina2023}, which is also associated with both water and class II methanol maser emission. The morphology of the 1.3~mm emission around MM1 resembles an extended, dense, dusty envelope and exhibits a slight elongation to the northeast.

Several other compact 1.3 mm continuum peaks (MM2, MM3, MM4, and MM5) are detected across the field (see Fig.~\ref{fig:cont}), indicating possible fragmentation of the parental molecular cloud. These sources are potential sites of low- to intermediate-mass star formation, though their evolutionary status requires verification through molecular line analysis and will be presented in the next paper.

The centimeter-wave sources VLA 1A, VLA 1B, and VLA 3, observed in \cite{Bayandina2023}, do not have clear compact counterparts in the 1.3 mm dust continuum. Instead, they are located along the periphery of the extended emission from the MM1 complex. Their positions coincide with the edges of a low-emission cavity and, crucially, with the brightest peaks of SiO emission (see Section \ref{sec:moleculas}). This spatial relationship strongly indicates that these sources trace shocked regions where a powerful outflow from the central MYSO (MM1/VLA 2) is interacting with the surrounding dense material, rather than being embedded protostars themselves. Notably, VLA 1A, the site of historical water maser super flares, is located at one of these prominent shock interfaces to the north of MM1. 

The source VLA 4 has no counterpart in the millimeter dust emission. This lack of associated dusty material suggests that VLA~4 is apparently not related to the G25.65+1.05 region and is likely a background extragalactic source.

To sum up, the overall structure of the continuum in the G25.65+1.05 region is characterized by a filamentary network surrounding the central source MM1. Some filaments may be potential accreting streamers that appear to be feeding the central massive envelope; however, this needs further investigation using molecular lines.

\subsection{Molecular line emission}
\label{sec:moleculas}

\begin{figure}
    \centering
    \includegraphics[width=0.9\linewidth]{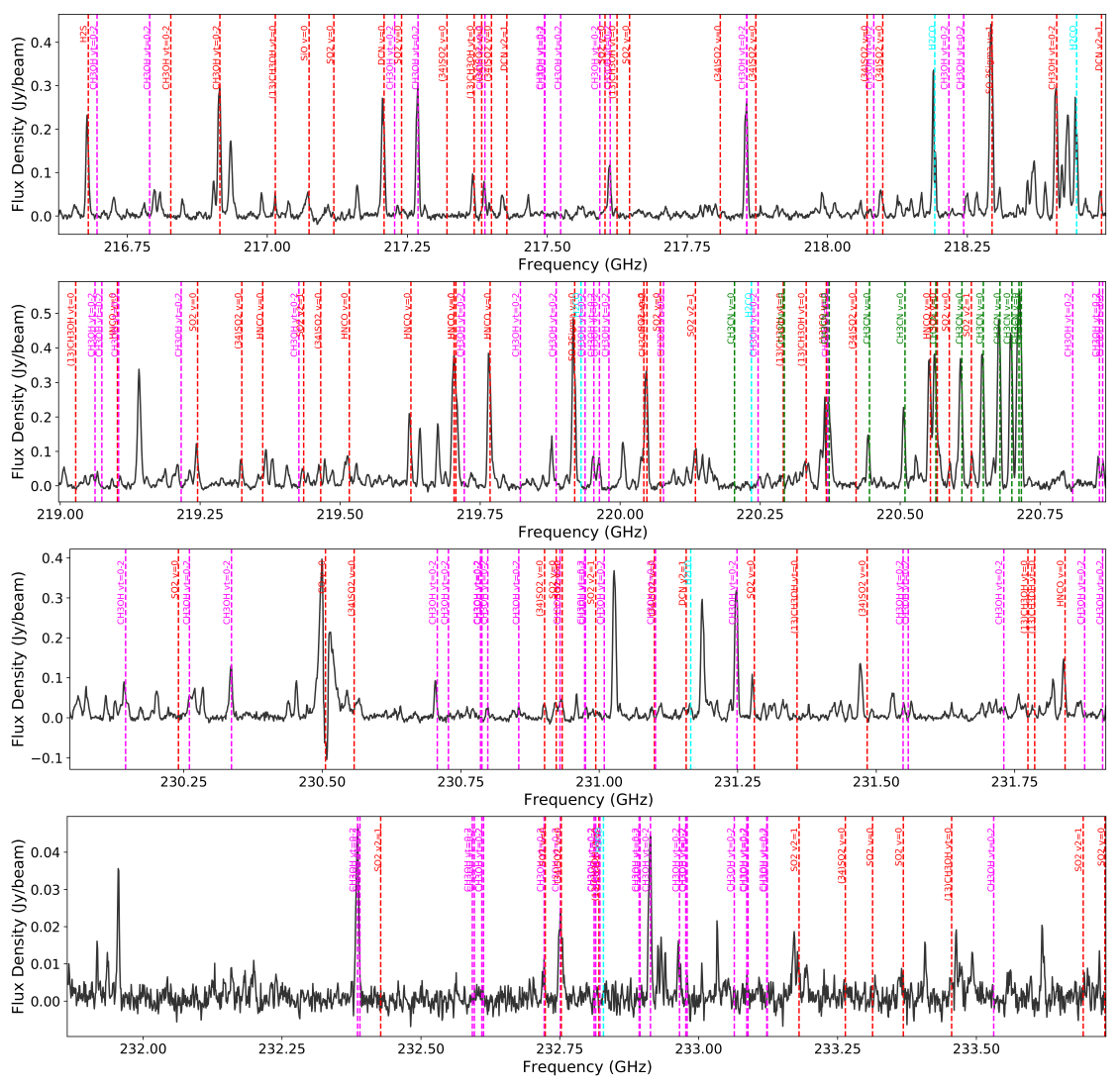}
    \caption{Spectra of the brightest source MM1 in the G25.65+1.05 region in four spectral windows. Vertical lines indicate molecular transitions of CH$_3$OH, SiO, CH$_3$CN, SO, DCN, and some other molecules.}
    \label{fig:VLA2_all_spw}
\end{figure}

The spectra of the MM1 source in four spectral windows are presented in Fig.~\ref{fig:VLA2_all_spw}. Molecular transitions of CH$_3$OH, SiO, CH$_3$CN, SO, DCN, and some others are indicated by vertical lines. These spectra are complex and contain a lot of spectral lines, which confirms that MM1 is a hot core. Previously, three molecules (CH$_3$CN, CH$_3$CCH, and CH$_3$OH) associated with the presence of a hot core were identified in \cite{2000A&A...361.1095D}. In our work, we observed rich molecular spectra of the source and identified a significantly larger number of molecules typical for hot cores that strongly supports the previously stated hypothesis.

\begin{figure}
    \centering
    \includegraphics[width=0.5\linewidth]{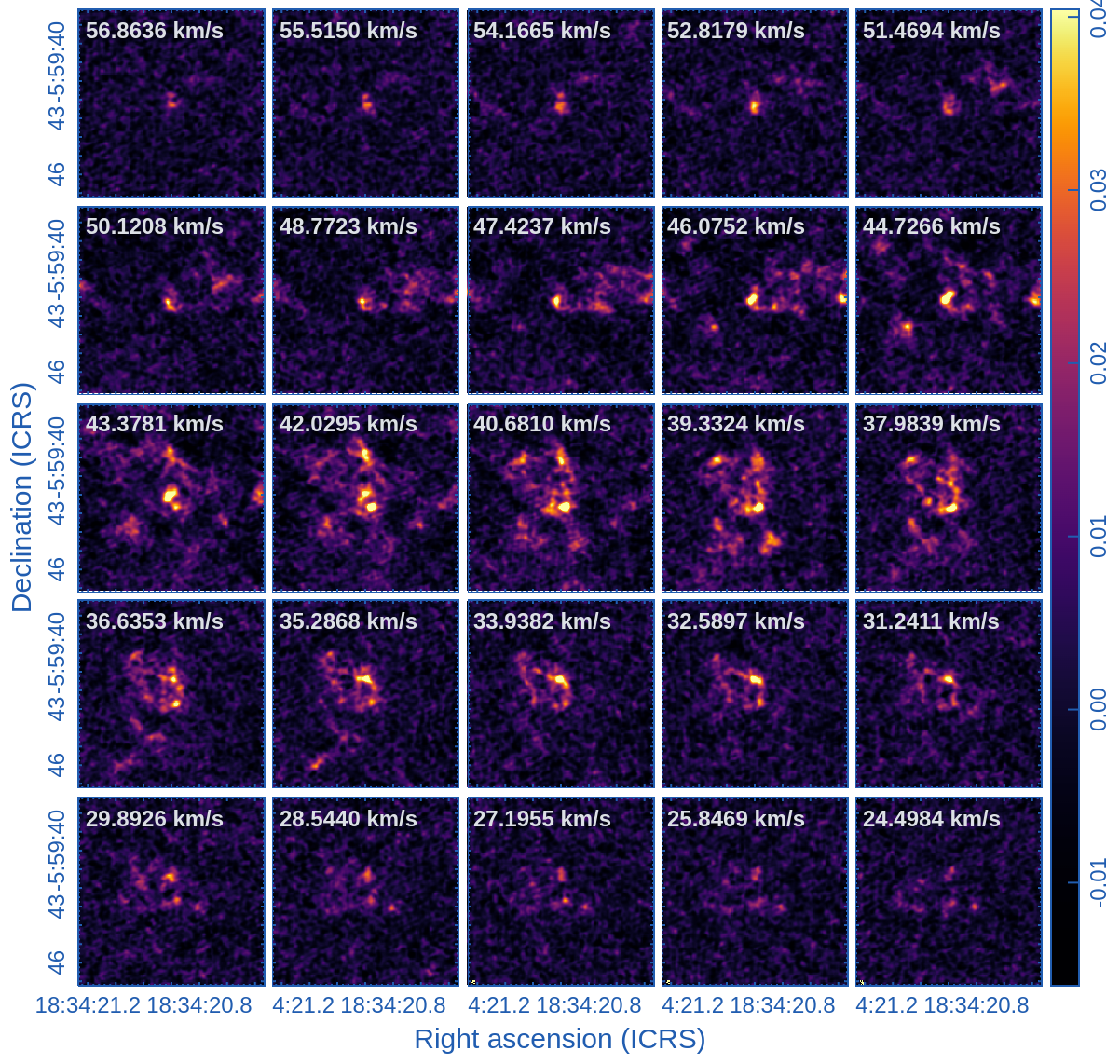}
    \caption{Spectral channels maps of the SiO (5-4) emission in the G25.65+1.05 region. The reference coordinates of the images are [RA, DEC]=[18:34:20.9, $-$05:59:42.2] and the image size is 10$^{''}\times$10$^{''}$. The velocity range corresponding to the channels is [24.50$-$58.86]~km/s.}
    \label{fig:sio_chan}
\end{figure}

\begin{figure}
\begin{minipage}{0.58\textwidth}
\includegraphics[width=0.99\linewidth]{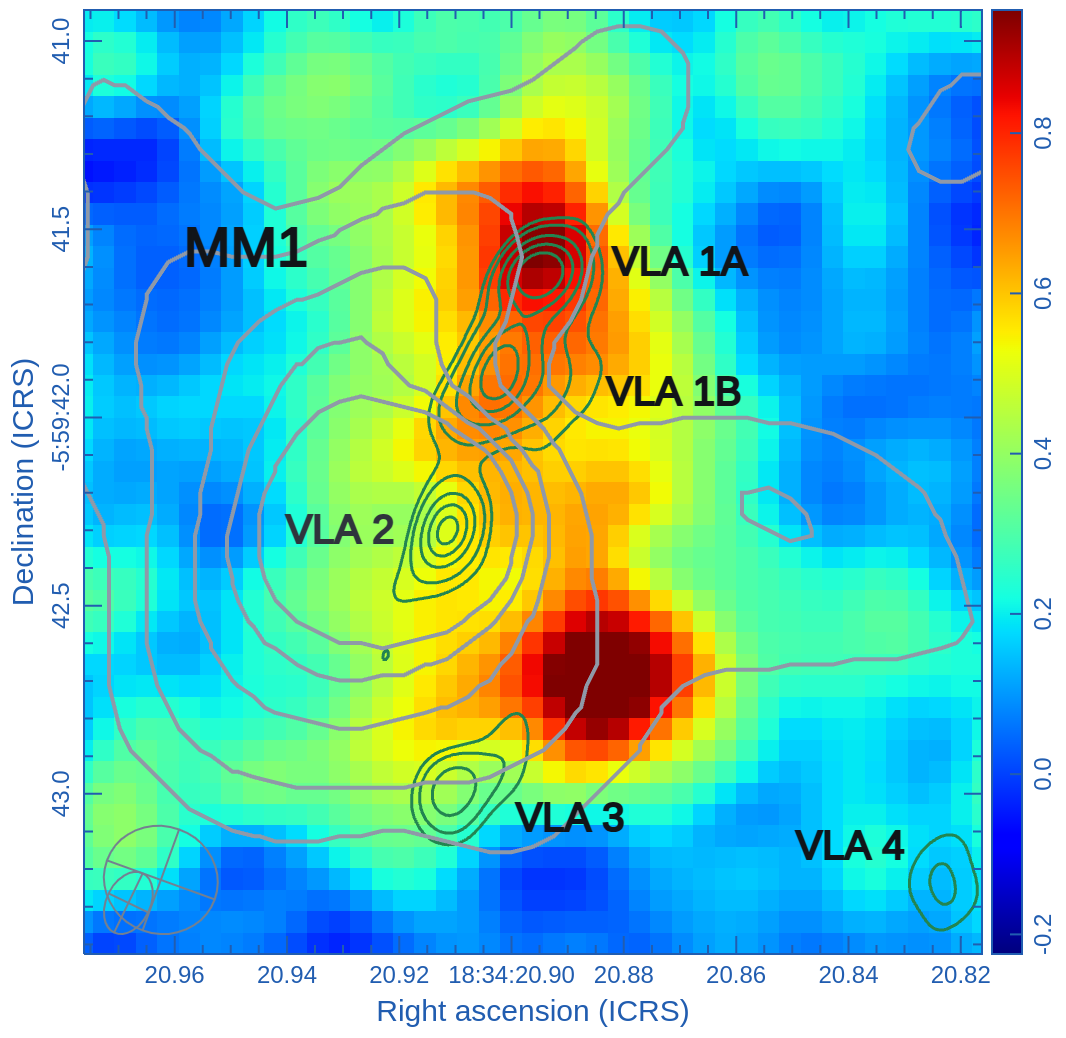}
\end{minipage}
\hfill
\begin{minipage}{0.41\textwidth}
\includegraphics[width=0.99\linewidth]{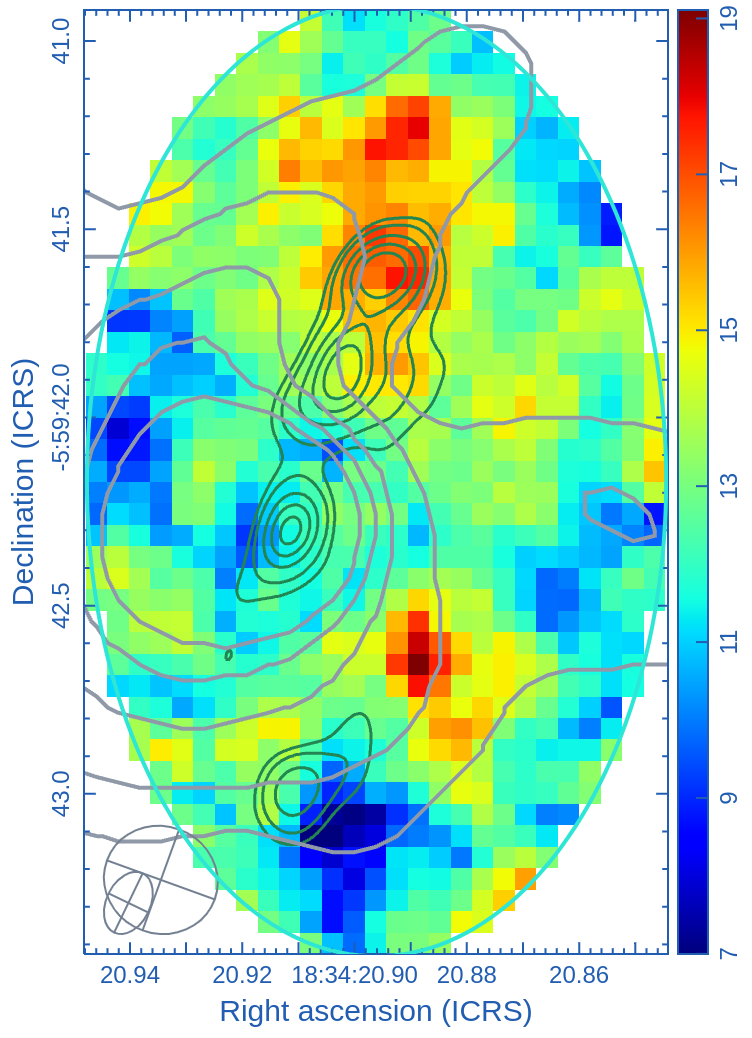}
\end{minipage}
\caption{{\bf Left panel}: Integrated (moment 0) map of the SiO (5-4) emission in the vicinity of the brightest millimeter source MM1 in the G25.65+1.05 region. Grey contours show 1.3-mm continuum revealed in ALMA observations in this work, the levels are set to (3, 5, 9, 13, 17)$\times\sigma$, where $\sigma=1.003\times10^{-3}$~Jy/beam. Green contours represent 2-cm continuum observed at the VLA in the work \cite{Bayandina2023}, the levels are set to (3, 5, 9, 13, 17)$\times\sigma$, where $\sigma=1.76\times10^{-5}$~Jy/beam.
{\bf Right panel}: Moment 2 map of SiO emission in the vicinity of main core MM1. The vertical colored panel indicates the velocity dispersion in the region. Grey contours indicate 1.3-mm continuum revealed in ALMA observations in this work, the levels are set to (3, 5, 9, 13, 17)$\times\sigma$, where $\sigma=1.003\times10^{-3}$~Jy/beam. Green contours show the continuum emission at 15 GHz from the work \cite{Bayandina2023}, the levels are set to (3, 5, 9, 13, 17)$\times\sigma$, where $\sigma=1.76\times10^{-5}$~Jy/beam.}
\label{fig:sio}
\end{figure}

\textbf{SiO emission.} Spatial distribution of the SiO (5-4) emission at 217.10~GHz is clearly seen in spectral channels corresponding to the velocity range [24.50$-$58.86]~km/s. Channel maps are presented in Fig. \ref{fig:sio_chan} and reveal an extended, complex structure with multiple patches of emission indicating a wide-angle outflow rather than a highly collimated jet.  
The integrated (moment~0) map of SiO (5-4) emission is presented in Fig. \ref{fig:sio}. The emission is mostly located to the northwest and southeast from the central continuum source MM1. The brightest SiO peaks show spatial coincidence with the centimeter-wave sources VLA 1A, VLA 1B, and VLA 3. Specifically, the northern SiO peak precisely aligns with VLA~1A, the site of the water maser super-flares. This spatial correlation strongly suggests that 15~GHz continuum VLA sources trace shocked regions where an outflow from the central MYSO impacts the surrounding dense material.
The moment 2 (velocity dispersion) map of SiO emission is presented in Fig.~\ref{fig:sio} and discussed in Section~\ref{sec:discussion}.

\textbf{CH$_3$CN emission.} Observations of methyl cyanide emission lines were used to probe dense gas and search for evidence of a potential disk structure. Previous studies have suggested the possible existence of such a disk based on near-infrared imaging of the source \cite{2002A&A...394..225Z}. 
Emission from both the lower-excitation CH$_3$CN (12-11) K=3 line (Fig.~5, left panel) and the higher-excitation K=7 lines (Fig.~5, right panel) are detected strongly and compactly towards MM1. The intensity-weighted velocity and velocity dispersion analyzed in the first and second moment maps of these lines (see Fig.~6) may indicate the presence of a possible disk.
The lower-excitation K=3 transition traces cooler outer regions and shows a distinct arc-like feature seen in the moment 1 map. Since lower-K lines are more likely optically thick, then we probably see only a part of the disk, as the rest of the emission is absorbed. The arc structure likely marks the disk edge where rotation is projected most strongly, while the opposite side is obscured by absorption. The higher-excitation K=7 moment 1 map displays a smoother velocity gradient without sharp arcs, consistent with a higher-K transition being more optically thin and tracing the full disk.
The moment 2 map of CH$_3$CN 12(3)-11(3) shows enhanced velocity dispersion on the side opposite to the K=3 moment 1 arc structure. This moment 2 feature likely traces the inner disk where Keplerian rotation changes rapidly with radius. The apparent anti-correlation between moment 1 and moment 2 arc positions for the K=3 transition may indicate an inclined Keplerian disk.

\begin{figure}
\begin{minipage}{0.49\textwidth}
\includegraphics[width=0.99\linewidth]{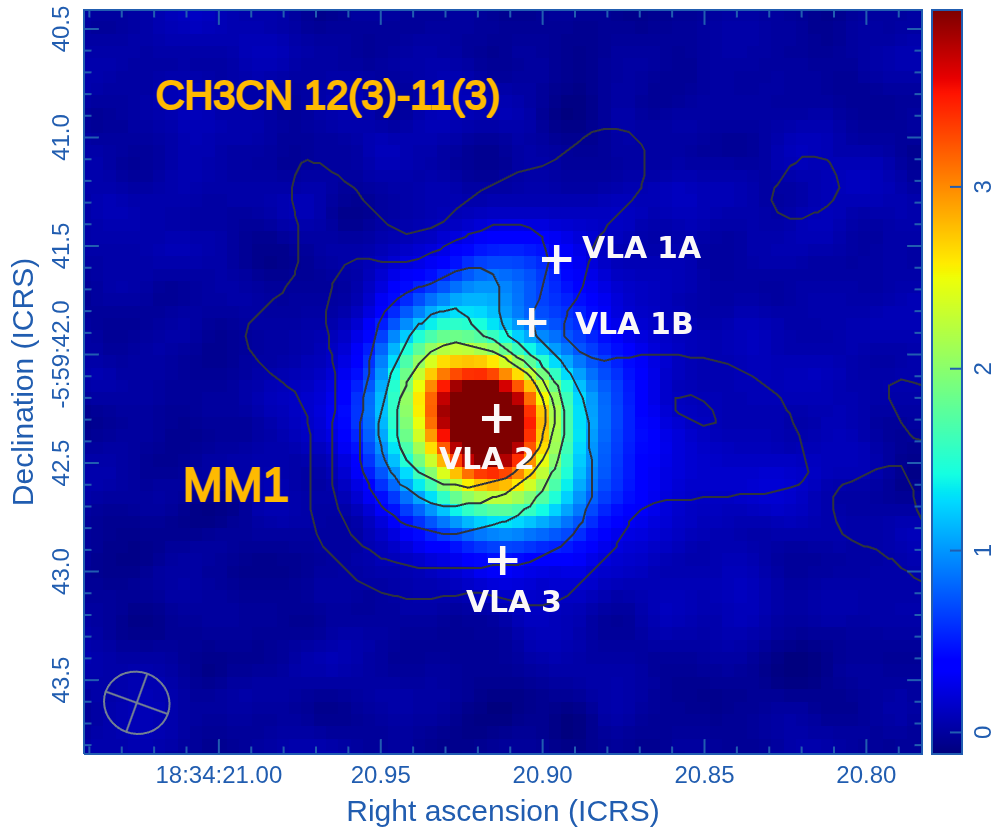}
\end{minipage}
\hfill
\begin{minipage}{0.49\textwidth}
\includegraphics[width=0.99\linewidth]{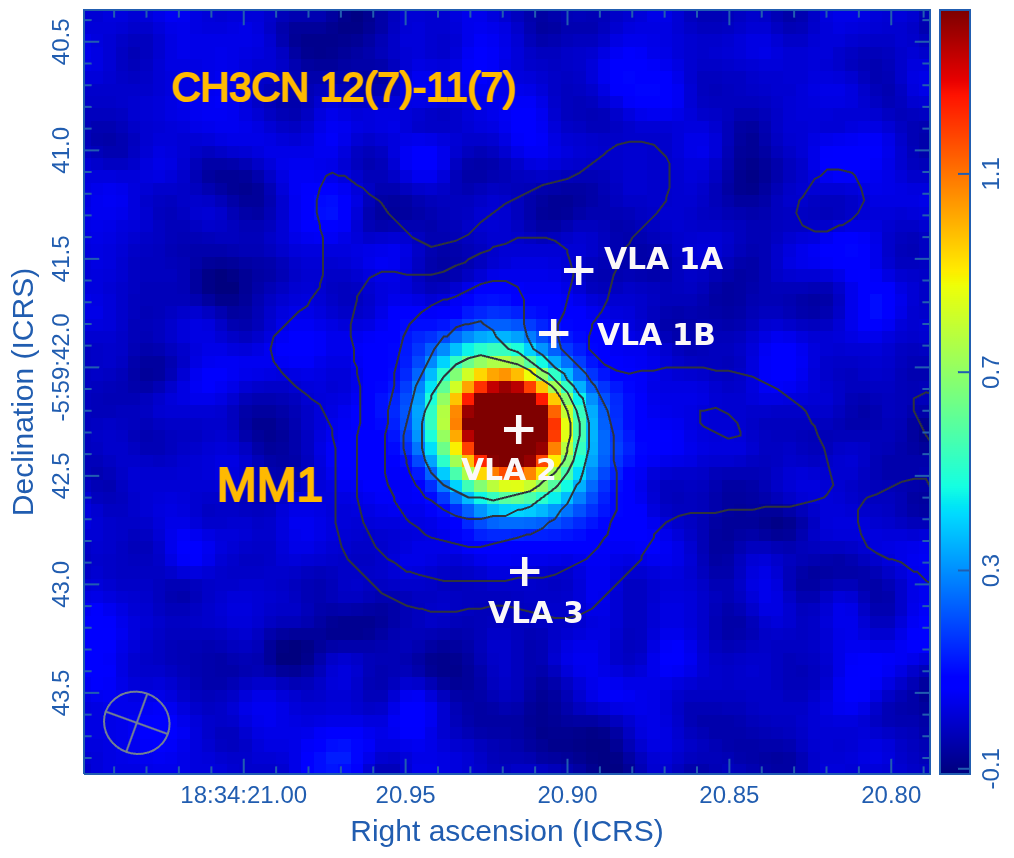}
\end{minipage}
\centering{\caption{Moment 0 maps of CH$_3$CN (12-11) K=3 and K=7 emission at 220.709 and 220.539 GHz, respectively, in the vicinity of the central source MM1. Black contours show 1.3-mm continuum revealed in ALMA observations in this work, the levels are set to (3, 5, 9, 13, 17)$\times\sigma$, where $\sigma=1.003\times10^{-3}$~Jy/beam. The size of the image is 4$^{''}\times$4$^{''}$ and the reference coordinates are [RA, DEC]=[18:34:20.9, $-$05:59:42.2]. The velocity range used for integration is [35.65$-$51.56]~km/s.}}
\label{fig:ch3cn}
\end{figure}

\begin{figure}
\begin{minipage}{0.45\textwidth}
\includegraphics[width=0.99\linewidth]{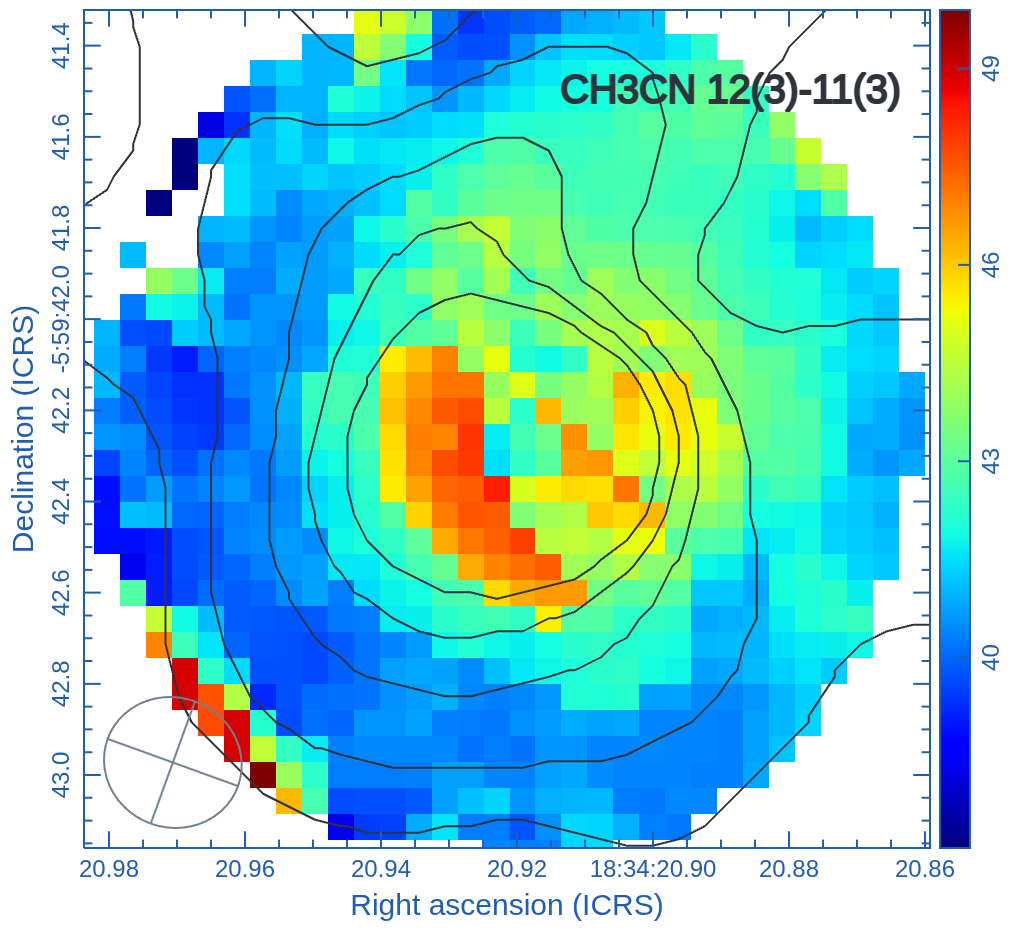}
\end{minipage}
\hfill
\begin{minipage}{0.45\textwidth}
\includegraphics[width=0.99\linewidth]{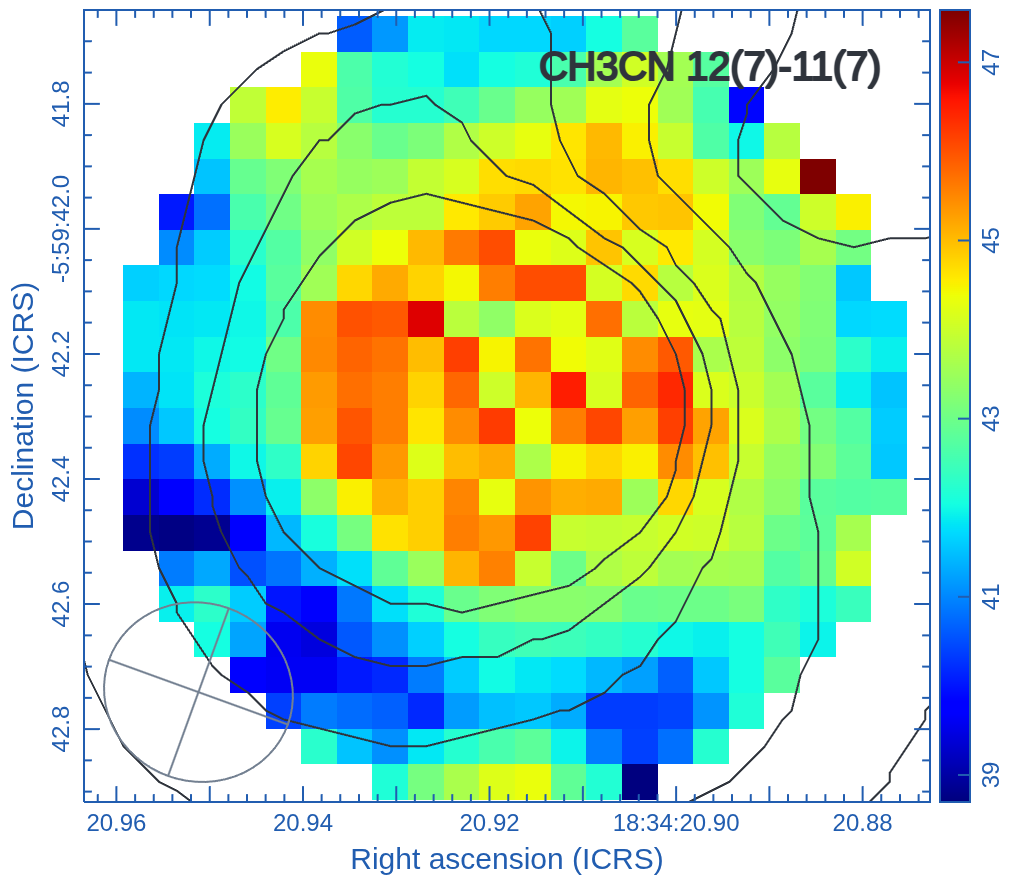}
\end{minipage}
\vfill
\begin{minipage}{0.45\textwidth}
\includegraphics[width=0.99\linewidth]{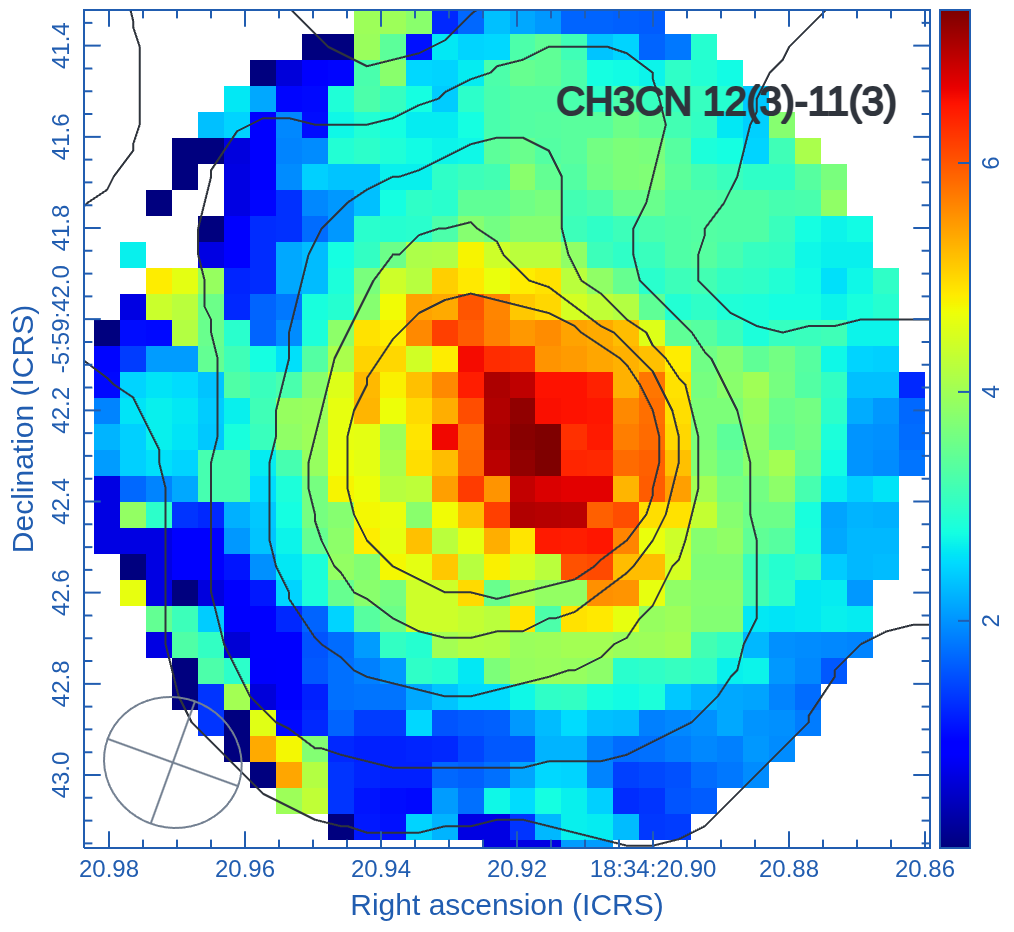}
\end{minipage}
\hfill
\begin{minipage}{0.45\textwidth}
\includegraphics[width=0.99\linewidth]{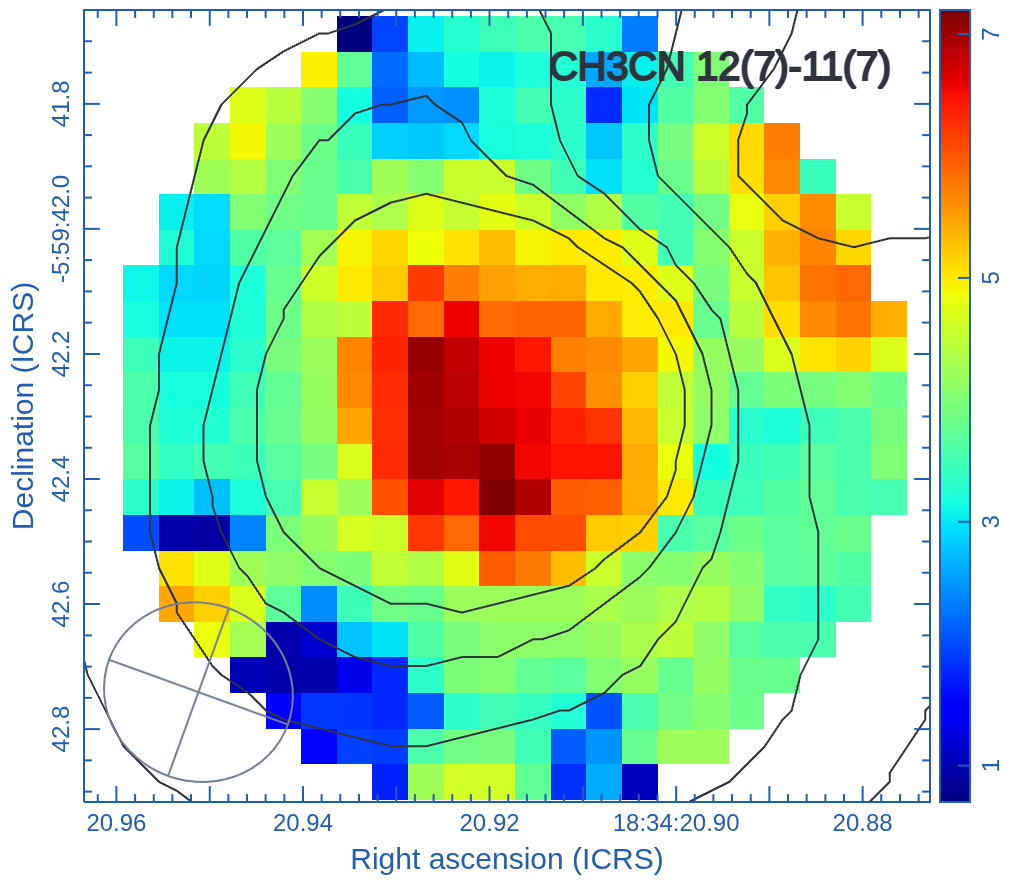}
\end{minipage}
\centering{\caption{Moment 1 (top) and moment 2 (bottom) maps of CH$_3$CN (12-11) K=3 and K=7 emission in the vicinity of the central source MM1. Black contours show 1.3-mm continuum revealed in this work, the levels are set to (3, 5, 9, 13, 17)$\times\sigma$, where $\sigma=1.003\times10^{-3}$~Jy/beam.}}
\label{fig:ch3cn_mom}
\end{figure}

\section{Discussion}
\label{sec:discussion}

\subsection{The overall morphology of the source}

The combined analysis of the molecular and continuum emission in G25.65+1.05 supports the interpretation that the central source MM1\,/\,VLA~2 is the driving source of the observed activity in the region. The presence of high-excitation line CH$_3$CN (K=7) confirms that MM1 is a hot molecular core. Two-dimensional Gaussian fitting of the MM1 continuum peak gives the angular size of 0.9$^{''}\times$0.8$^{''}$. Assuming the distance to the region of 2.08 kpc \cite{2016ApJ...823...77R}, it gives the linear size of the source of about 2000 AU, which is a typical value for the emitting dust shells of massive protostars. Thus, our observations suggest a distance of about 2~kpc rather than the larger distance of 12.5 kpc suggested in \cite{{2011MNRAS.417.2500G}}. The lack of strong molecular line emission towards other millimeter continuum peaks (MM2, MM8, MM10) suggests they may contain less evolved protostars or starless cores (this point will be investigated in further work).

Spatial distribution of SiO (5-4) emission shows regions of interaction between the outflow produced by the central source MM1 and surrounding material. The overall morphology of the SiO emission patches seen in $\sim$25 spectral channels (see Fig.~\ref{fig:sio_chan}) is complex and suggests that the outflow has rather a wide angle, or even multiple jets may be present in the region. We propose that SiO emission traces the areas where the outflow breaks through the inhomogeneous envelope surrounding the MM1 source. As seen from the continuum map, the close environment of the MM1 peak is brighter to the east from the MYSO, which means that the eastern parts of the dust envelope are denser. Therefore, it is easier for the outflow to break through in the northwest and southwest directions, that is, where dips in the continuum are visible (see Fig.~\ref{fig:cont}). Thus, the powerful wide-angle outflow driven by MM1 creates shocked regions at the interfaces with the ambient cloud, which are observed as strong SiO emission and are coincident with the non-thermal centimeter-wave sources VLA~1A, 1B, and 3. It is likely that the H$_2$O masers, revealed earlier in the G25.65+1.05 region and associated mainly with the VLA~1A source, are located at the edge of the cavity formed by the outflow. 
In general, the overall structure of the 1.3 mm continuum, SiO maps, and VLA continuum hint at the presence of multiple jets in the system.

\subsection{Molecular emission in the region of water maser super flares}

In order to clarify the nature of the VLA~1A and VLA~1B sources, especially the former, which is lying behind the H$_2$O maser super flares, we extracted molecular spectra corresponding to these two sources. The masks used for the extraction corresponded to the 5 sigma level of centimeter continuum emission at 15~GHz (see Fig.~7, left panel).

\begin{figure}
\begin{minipage}{0.37\textwidth}
\includegraphics[width=0.99\linewidth]{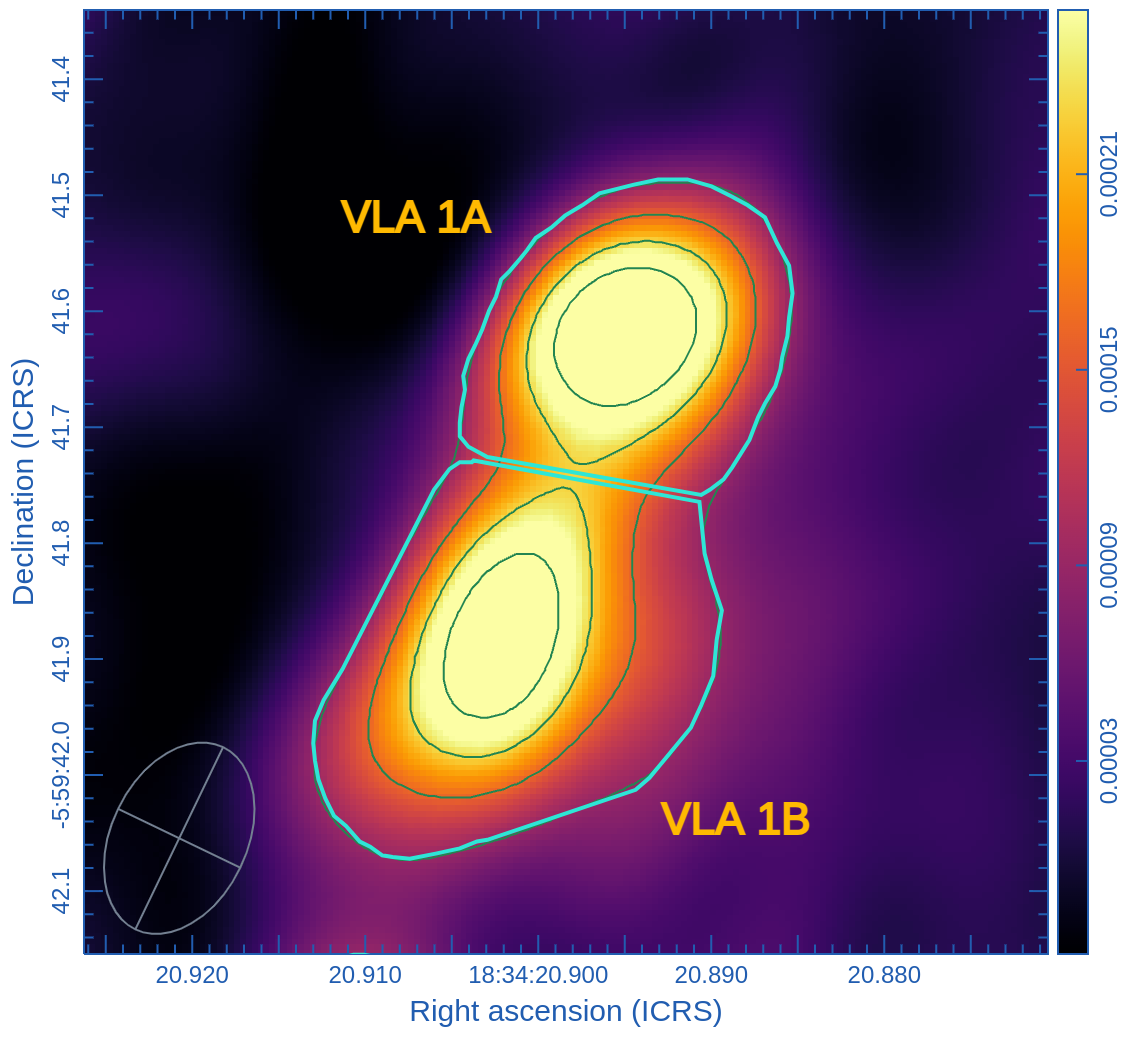}
\end{minipage}
\hfill
\begin{minipage}{0.58\textwidth}
\includegraphics[width=0.99\linewidth]{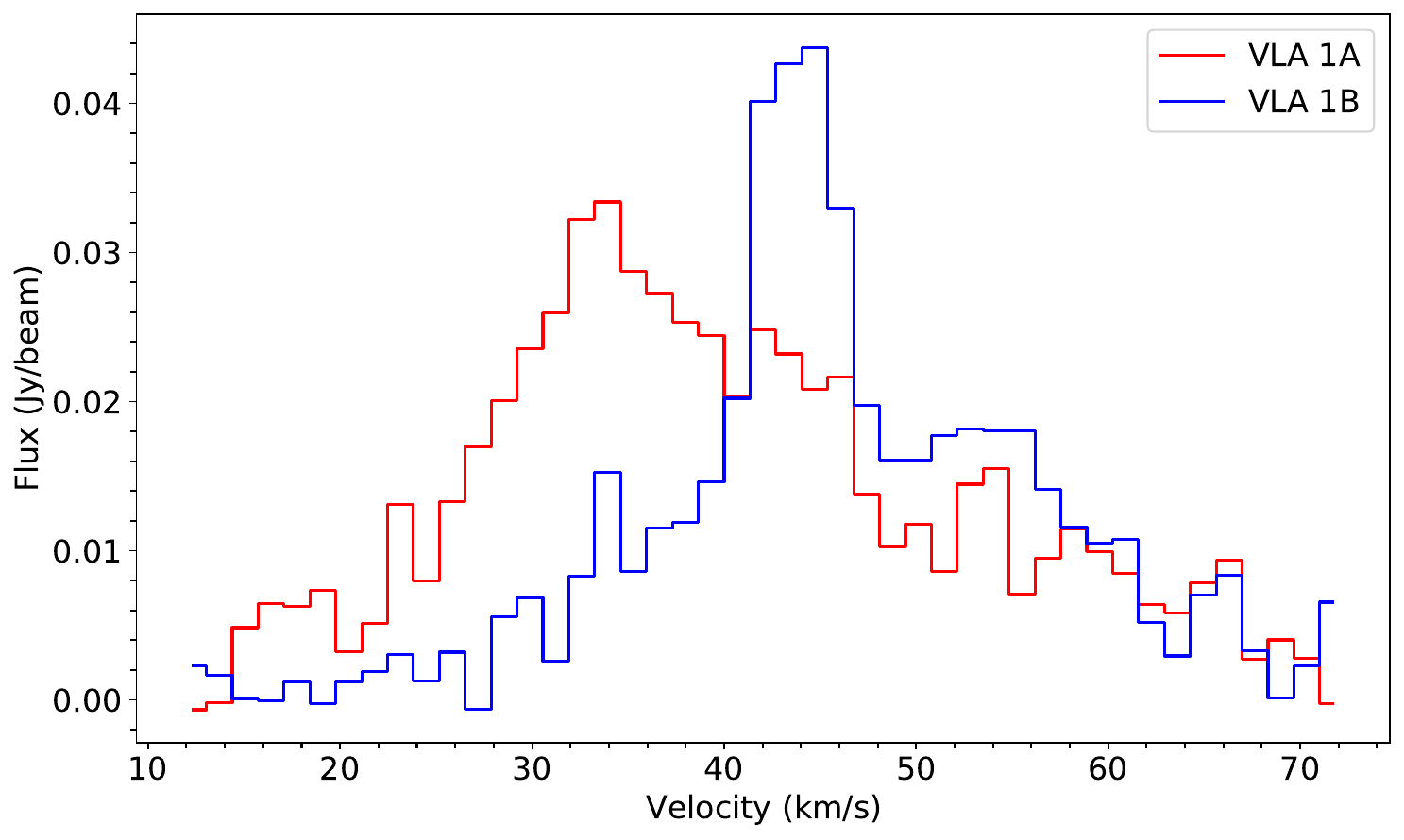}
\end{minipage}
\centering{\caption{\textit{Left panel}. Masks (cyan curves) used for extraction of molecular emission in the sources VLA 1A and VLA 1B. The background image is the 15 GHz continuum from the work \cite{Bayandina2023}. \textit{Right panel}. Spectral profiles of SiO (5-4) emission in the VLA 1A (red) and VLA 1B regions (blue).}}
\label{fig:masks_sio_profile}
\end{figure}

\begin{figure}
    \centering
    \includegraphics[width=0.9\linewidth]{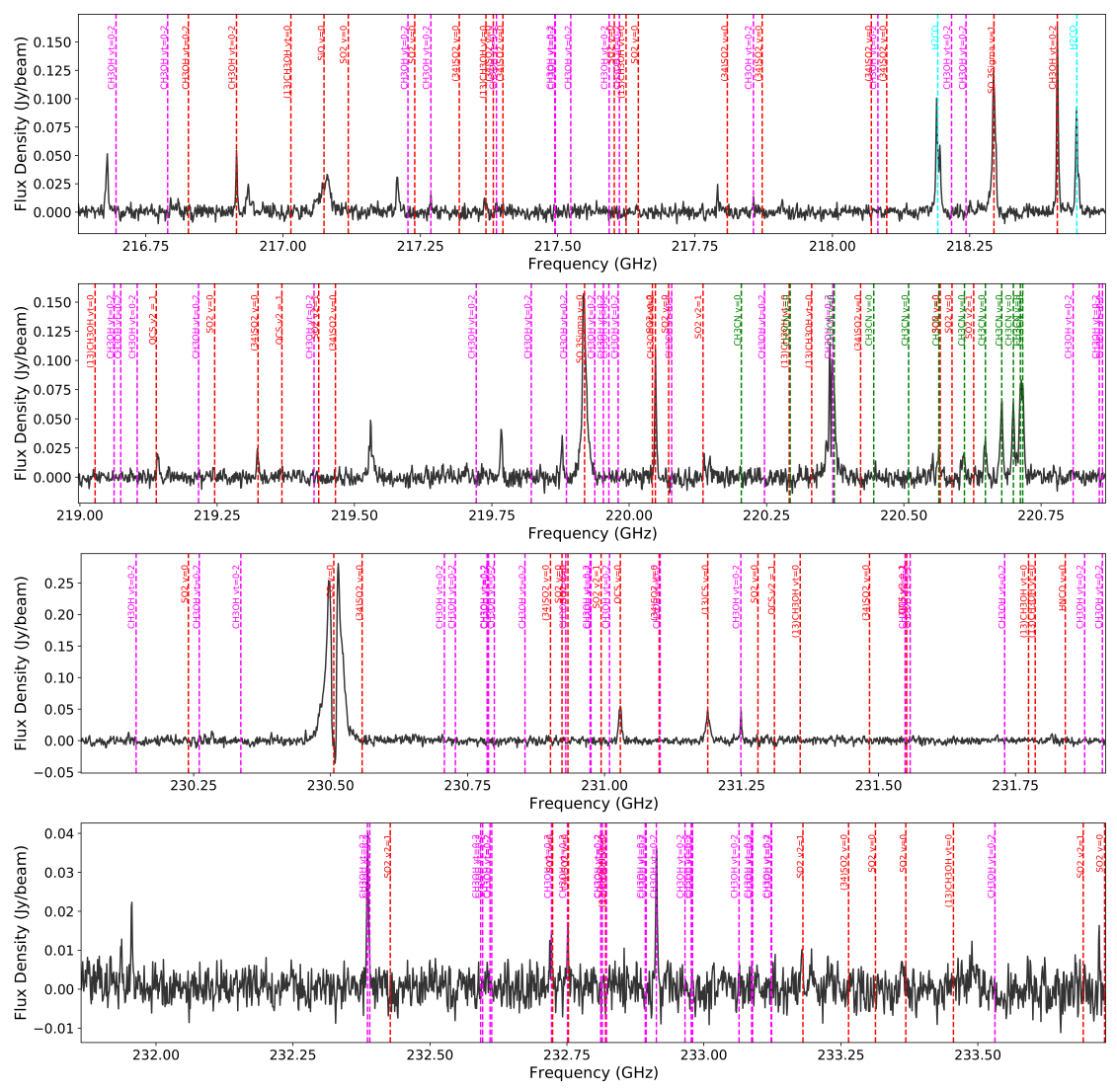}
    \caption{Molecular emission in the VLA 1A source associated with H$_2$O maser super flares in the region G25.65+1.05 in all observed spectral windows.}
    \label{fig:VLA1A_all_spw}
\end{figure}

\begin{figure}
    \centering
    \includegraphics[width=0.9\linewidth]{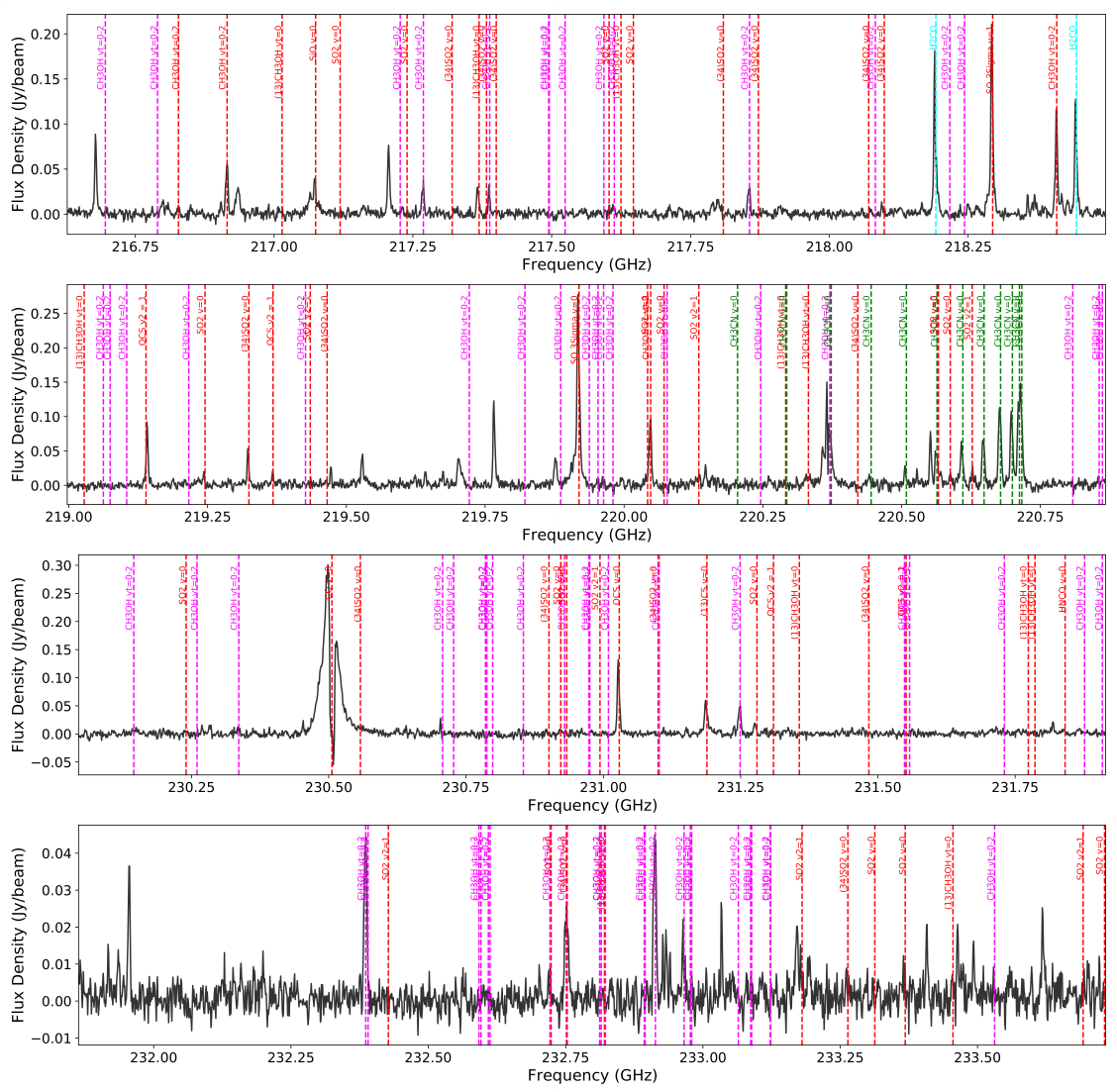}
    \caption{Molecular emission in the VLA 1B source in the region G25.65+1.05 in all observed spectral windows.}
    \label{fig:VLA1B_all_spw}
\end{figure}

Figures \ref{fig:VLA1A_all_spw}-\ref{fig:VLA1B_all_spw} present the molecular spectra of the VLA~1A and VLA~1B sources across the four observed spectral windows (216.7–218.7, 219–221, 230–232, and 232–234 GHz). Vertical dashed lines mark transitions of various molecules at a systemic velocity of V$_{LSR}=41.7$~km/s. The spectra contain lines characteristic of regions of intense interaction between outflows and surrounding material, such as SiO, CS, and SO, as well as tracers of dense gas H$_2$S and DCN. Notably, the SiO and CS lines are broader in VLA~1A than in VLA~1B, indicating a more turbulent environment. This is demonstrated by the SiO line profiles in Fig.~7 (right panel), where the emission from VLA~1A is significantly broader and exhibits a more complex structure than that of VLA~1B. The moment~2 velocity dispersion map of SiO emission clearly shows the peak exactly at the position of the VLA~1A source (see Fig. \ref{fig:sio}). A larger second moment indicates more developed chaotic (turbulent) movements. This higher level of turbulence directly increases the probability of transient events in VLA~1A. Specifically, it favors the short-term alignment of regions with population inversion along the line of sight, which is a prerequisite for strong maser amplification. Therefore, the observed turbulent conditions in VLA~1A provide a direct explanation for the origin of the maser super flares associated with this particular source.

\section{Conclusions}
\label{sec:conclusion}

The analysis of high-resolution ALMA observations of the dust continuum and molecular lines toward the massive star-forming region G25.65+1.05 is presented with a focus on elucidating the nature of the continuum sources previously identified in VLA observations.

The main results are summarized as follows:

1. The bright 1.3 mm continuum source MM1 is identified as the central massive young stellar object in the center of a fragmented, filamentary cloud, confirmed by its precise spatial coincidence with the centimeter-wave source VLA~2. The linear size of the MM1 source is about 2000 AU at a distance of 2.08 kpc. The observed complex molecular spectra of the MM1 source, especially the presence of high-excitation CH$_3$CN (12-11) K=7 emission, confirms that MM1 is a hot molecular core. CH$_3$CN lines reveal kinematic signatures consistent with a possible rotating disk structure in MM1.

2. The centimeter continuum sources VLA~1A, VLA~1B, and VLA~3 are not associated with compact millimeter continuum counterparts. Instead, their positions coincide with the brightest peaks of SiO emission at the periphery of the MM1 complex. This spatial relationship strongly indicates that these sources trace shocked regions where a wide-angle outflow from the central MYSO MM1 interacts with the surrounding dense material, rather than being embedded protostars themselves. The source VLA~4 is lacking any millimeter counterpart and may be a foreground extragalactic object.

3. The compact continuum source VLA~1A, associated with extreme H$_2$O maser activity, is confirmed to be a shock interface rather than a protostar. This clarification strongly supports the previously proposed model in which maser amplification results from the overlap of masering regions within an outflow-driven shock.
The observed molecular emission demonstrates that the VLA~1A source resides in a highly turbulent environment, a characteristic not seen in VLA~1B. This turbulence directly promotes the transient conditions needed for maser action, providing a conclusive link between the nature of VLA~1A and the origin of the H$_2$O maser super flares.

\section*{Acknowledgements}

The work was supported by the Ministry of Science and Higher Education of the Russian Federation, project no. 075-15-2024-538.

Authors are grateful to O. S. Bayandina for useful comments and for providing the original centimeter continuum data used in this work for comparison with our ALMA data.

This paper makes use of the following ALMA data: ADS/JAO.ALMA 2021.1.00311.S. ALMA is a partnership of ESO (representing its member states), NSF (USA) and NINS (Japan), together with NRC (Canada), NSTC and ASIAA (Taiwan), and KASI (Republic of Korea), in cooperation with the Republic of Chile. The Joint ALMA Observatory is operated by ESO, AUI/NRAO and NAOJ.

\section*{References}
\bibliography{refs} 


\end{document}